\documentclass[aps,prl,twocolumn,superscriptaddress]{revtex4-2}
\usepackage{comment}
\usepackage[utf8]{inputenc}
\usepackage[T1]{fontenc}
\usepackage{lmodern}
\usepackage{graphicx}
\usepackage{xparse}
\usepackage{amsmath}
\usepackage{amssymb}
\usepackage{microtype}
\usepackage{braket}
\usepackage{etoolbox}

\usepackage{verbatim}

\usepackage{tikz}
\usetikzlibrary{calc, arrows.meta}
\newcommand{\tikzmark}[1]{\tikz[overlay,remember picture] \node (#1) {};}

\usepackage[breaklinks=true,colorlinks, citecolor=blue]{hyperref}

\newcommand{\cexp}[1]{\langle #1 \rangle}
\newcommand{\e}[0]{{\mathrm{e}}}

\let\oldparagraph\paragraph
\RenewDocumentCommand{\paragraph}{s o m}{%
  \IfBooleanTF{#1} % if starred version is used
    {%
      \IfNoValueTF{#2}% if no optional argument is given
        {\oldparagraph*{\textbf{#3}}}%
        {\oldparagraph*[\textbf{#2}]{\textbf{#3}}}%
    }%
    {%
      \IfNoValueTF{#2}% if no optional argument is given
        {\oldparagraph{\textbf{#3}}}%
        {\oldparagraph[\textbf{#2}]{\textbf{#3}}}%
    }%
}

\begin{document}

\title{Chiral quantum state circulation from photon lattice topology}

\author{Souvik Bandyopadhyay}
\affiliation{Department of Physics, Boston University, 590 Commonwealth Avenue, Boston, Massachusetts 02215, USA}

\author{Anushya Chandran}
\affiliation{Department of Physics, Boston University, 590 Commonwealth Avenue, Boston, Massachusetts 02215, USA}

\author{Philip JD Crowley}
\affiliation{Department of Physics, Harvard University, Cambridge, Massachusetts 02138, USA}
\affiliation{Department of Physics and Astronomy, Michigan State University, East Lansing, Michigan 48824, USA}
\email{philip.jd.crowley@gmail.com}

\date{\today}

\begin{abstract}
Chiral quantum state circulation is the unidirectional transfer of a quantum state from one subsystem to the next. It is essential to the working of a quantum computer; for instance, for state preparation and isolation.
We propose a cavity-QED architecture consisting of three cavities coupled to a qubit, in which \emph{any} photonic state of cavity 1 with sufficiently many photons circulates to cavity 2 after a fixed time interval, and then to cavity 3 and back to 1. 
Cavity-state circulation arises from topologically protected chiral boundary states in the associated photon lattice and is thus robust to perturbation.  
We compute the circulation period in the semi-classical limit, demonstrate that circulation persists for time-scales diverging with the total photon number, and provide a Floquet protocol to engineer the desired Hamiltonian. 
Superconducting qubits offer an ideal platform to build and test these devices in the near term. 
\end{abstract}

\maketitle

Achieving large-scale, fault-tolerant quantum information processing is a central goal of modern physics~\cite{Feynman1982Simulating,preskill2018quantum}, with road-maps projecting systems of $\approx 10^5$ qubits within the coming decade~\cite{GoogleQuantumAI_Roadmap,IBM_roadmap,gidney2025factor}. 
There are several challenges; in addition to the well-known challenge of logical error detection and correction~\cite{shor1995scheme,campbell2017roads,terhal2015quantum}, the precision and speed of control at the physical qubit scale needs to increase~\cite{gidney2025factor,gouzien2021factoring,divincenzo2000physical,jeffrey2014fast}. 
How these challenges will be overcome remains to be seen~\cite{preskill2023quantum}. Meanwhile, progress can be made by refining existing protocols and devising new techniques to rapidly manipulate qubits. 

We focus on a particular bottleneck: the rapid readout of a cavity qubit state, and reset to the vacuum state. 
This problem is particularly acute for superconducting processors that encode quantum information in high-Q microwave cavities~\cite{geerlings2013demonstrating,jeffrey2014fast,chen2023transmon,bengtsson2024model,maurya2024demand,xiao2025flexible,ding2025multipurpose}. 
By design, high-Q cavities are well isolated, and thus difficult to externally tune or couple to. 
If the cavity state is known, it can be reset rapidly using linear (Gaussian) operations. 
However, resetting an unknown state requires dissipative and nonlinear interactions that, in a high-Q cavity, are intrinsically slow~\cite{pietikainen2024strategies,ding2025multipurpose}.
Similar challenges apply to rapid readout, with the additional requirement of unidirectional transmission: the measurement outcome must be transmitted without thermal noise flowing back in from the control/readout lines.

A key ingredient in our approach is the intrinsic topological response of few-body photonic systems. 
Recent work has shown that topologically protected responses can be harnessed to unidirectionally pump photons between cavities~\cite{martin2017topological,peng2018time,kolodrubetz2018topological,crowley2019topological,nathan2019topological,crowley2020half,nathan2020quantum,long2021nonadiabatic,boyers2020exploring,long2024topological,wu2025geometric}, and engineer photon current loops between multiple cavities in a driven cavity-qubit system~\cite{yuan2018synthetic,Deng:2022aa,cai2021topological,zhang2025synthetic,yuan2024quantum}. 

\begin{figure}[tb]
    \includegraphics[width=1.01\columnwidth,height=7.5cm]{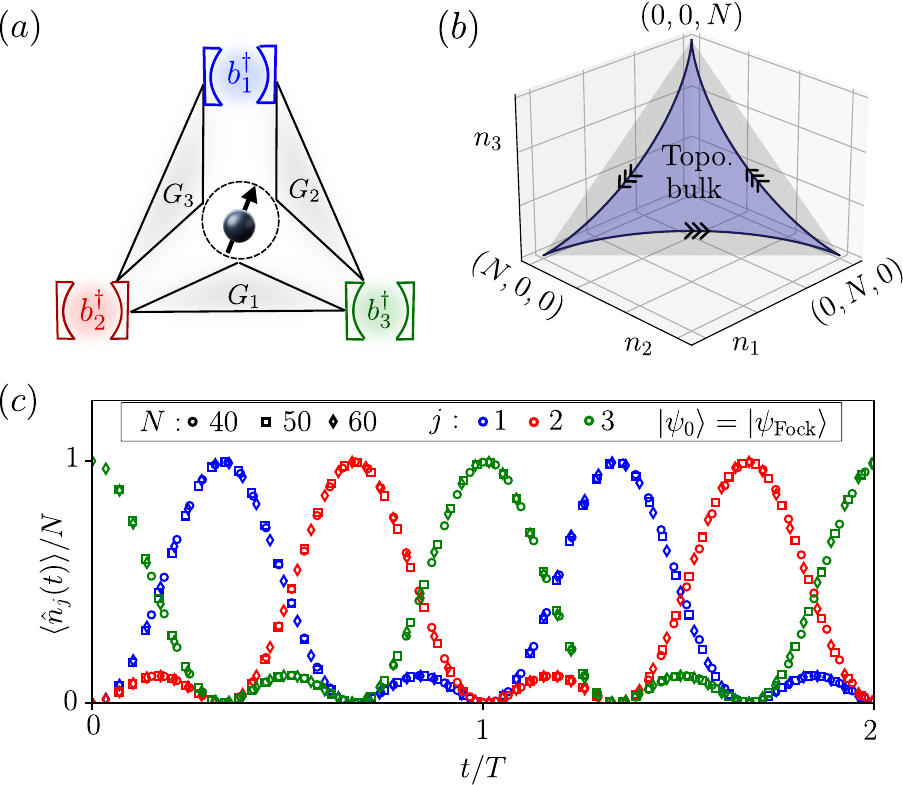}
	\caption{\emph{Topological photon circulation}: (a) Three cavities $b_j^\dagger$ interact with a qubit via three-body couplings $G_j$ (b) The Hamiltonian induces hopping on a triangular \emph{photon lattice} on the Fock space plane $\sum_j n_j = N$ (gray). This model has topological boundary modes which carry photon current (black arrows). (c) The boundary modes result in circulation~\eqref{eq:circulator} of photons between the cavities, as seen in the dynamics of cavity populations $\cexp{\hat{n}_j}$, shown here for an initial Fock state~\eqref{eq:initial_states}.}
	\label{fig:p1}
\end{figure}

Using this topological response, we propose a few-body (three-cavity, one-qubit) device which exhibits the high-fidelity, unidirectional quantum state transfer necessary for rapid and robust cavity reset/readout. 
The model hosts chiral circulating probability currents in Fock space (see Fig.~\ref{fig:p1}), causing a product state of the qubit and cavity modes to undergo \emph{quantum state circulation} with period $T$:
{
\begin{align}
        \tikzmark{st1start}|\psi,0, 0\rangle |\sigma_1\rangle\tikzmark{st1end} \xrightarrow[]{T/3} |0,\psi,0\rangle |\sigma_2\rangle \xrightarrow[]{T/3} \tikzmark{st3start}|0,0,\psi\rangle |\sigma_3\rangle \tikzmark{st3end}
	\label{eq:circulator}
\end{align}}%{Stealth[scale=1.2]}
\begin{tikzpicture}[overlay, remember picture]
  \coordinate (st1mid) at ($(st1start)!0.5!(st1end)$);
  \coordinate (st3mid) at ($(st3start)!0.5!(st3end)$);
  \coordinate (below st3mid) at ($(st3mid)+(0,-2.0ex)$);
  \coordinate (below st1mid) at ($(st1mid)+(0,-2.0ex)$);
  \draw[->={latex},line width=0.1pt] ($(st3mid)+(0,-0.5ex)$) -- (below st3mid) -- (below st1mid) -- ($(st1mid)+(0,-0.5ex)$)
    node[midway, below, yshift=-1.0ex,xshift=18ex] {\scriptsize $T/3$};%\vspace{6cm}
\end{tikzpicture}
\newline
for \emph{any} single cavity state $\ket{\psi}$ of sufficiently many photons, and specific qubit states $\ket{\sigma_i}$ for $i=1,2,3$. Significantly, this resets the state of the first cavity $\ket{\psi}$ to the vacuum $\ket{0}$ in time $T/3$, moving $\ket{\psi}$ to the second cavity where it may be read out and reset while operations on the first cavity continue.
Similarly, running the process in reverse loads any state into the cavity in the same time.
By choosing the Q-factors of the 3 cavities as desired, the ability to shuttle arbitrary states rapidly between cavities reconciles the competing demands: high-Q for storage, but not for fast reset and readout.

We demonstrate that the quantum state circulation~\eqref{eq:circulator} is high-fidelity and topologically robust, persisting under detuning of the cavity frequencies and generic perturbations of the Hamiltonian.
Quantitatively, the imprecision for the state transfer in one cycle is $\sim1/\sqrt{N}$ (for $N$ photons), so that the state completes $\sim\sqrt{N}$ circulations before degrading. The state circulation follows from the presence of chiral boundary modes in Fock space (see Fig.~\ref{fig:p1}(b)). 
Per the bulk-boundary correspondence~\cite{bansil2016colloquium}, these boundary modes are protected by the topology (specifically, the band Chern numbers) of the bulk Hamiltonian in Fock space. 
Consequently, we refer to the Hamiltonian in the Fock basis as acting on a \emph{photon lattice} to emphasize this correspondence with topological band theory.

The model Hamiltonian we propose includes intrinsic three-body interactions. However, we show it may be obtained via standard Floquet engineering from two-body interactions. 
This cavity-photon circulation is thus realizable in state-of-the-art cavity-QED experiments.

\paragraph*{Model and photon lattice:} The setup consists of three cavities and a qubit, with the qubit mediating hopping of photons between cavities (Fig~\ref{fig:p1}(a)). The Hamiltonian is,
\begin{equation}\label{eq:ham}
	H = \Delta\sigma_z+\omega\hat{N}+ \left(b_1^{\dagger}b_{2}G_{3}+b_2^{\dagger}b_{3}G_{1}+b_3^{\dagger}b_{1}G_{2}+\text{h.c.}\right).
\end{equation}
Here $\hat{N}=\sum_j \hat{n}_j$ where $\hat{n}_j=b_j^{\dagger}b_j$ counts the photons in the $j$th cavity; the qubit splitting is $2\Delta$; and the qubit operators are defined as $G_j= i g(\cos(2\pi j/3)\sigma_x+\sin(2\pi j/3)\sigma_y+ i \sigma_z)$, where $\sigma_\alpha$ are the usual Pauli matrices. All three cavities have the same natural frequency $\omega$. We write hats on the operators $\hat{N}$ and $\hat{n}_j$ to avoid confusion with their corresponding eigenvalues ($N$ and $n_j$).

The Hamiltonian has several symmetries: (i) It conserves total photon number $[H,\hat{N}] = 0$. (ii) It breaks time reversal symmetry with $\mathcal{T}=\mathcal{K}$, where $\mathcal{K}$ denotes complex conjugation. (iii) $H-\omega \hat{N}$ anti-commutes with the anti-unitary operator $\mathcal{P}=i \sigma_y \mathcal{K}$. (iv) $H$ has a three-fold rotational symmetry under the unitary $U_{\mathrm{C}_3}$ which cyclically permutes the cavity indices and rotates the qubit by $2\pi/3$ about the $z$-axis: $U_{\mathrm{C}_3} b_j U_{\mathrm{C}_3}^\dagger = b_{j+1}$, $U_{\mathrm{C}_3} G_j U_{\mathrm{C}_3}^\dagger = G_{j+1}$. These symmetries have immediate consequences. By (i) the dynamics is restricted to fixed total $N$ planes in Fock space, while (ii) allows for stationary states with photon currents related by $\mathcal{P}$ anti-symmetry (iv). Property (iii) implies that within a number sector the energy spectrum is symmetric about $E=\omega N$.   
    
The Hamiltonian can be represented as an inhomogeneous tight-binding model in Fock space~\cite{crowley2019topological,sambe1973steady,ho1983semiclassical,icfo24,Yuan18,saugmann2023fock}. Consider the three-dimensional \emph{photon lattice}, with sites $\vec{n} = (n_1,n_2,n_3)$, and two orbitals per site corresponding to the two qubit configurations, with corresponding basis vectors $\ket{\vec{n}, \sigma_z=\pm 1}$. Eq.~\eqref{eq:ham} produces a tight-binding model on this lattice, in which the natural frequencies of the cavities induce a uniform electric field $\vec{\omega}=\omega(1,1,1)$. The first term in~\eqref{eq:ham} produces a uniform on-site energy splitting; the second term encodes the electric field $\vec{\omega}$; while the third term induces hopping amplitudes between neighboring sites of the same total photon number $N$. The hopping amplitudes are site (or $\vec{n}$) dependent due to the Bose enhancement; this inhomogeneity is crucial to high-fidelity circulation, as we discuss later. Eq.~\eqref{eq:ham} produces a two-dimensional nearest-neighbor hopping model on a triangular lattice lying in the plane perpendicular to $\vec{\omega}$. This lattice is finite (see Fig.~\ref{fig:p1}(b)), as follows from $n_j\geq 0$,. Lastly, we note the symmetries of $H$ are reflected in the tight-binding model. In particular, $U_{\mathrm{C}_3}$ symmetry appears as a composite symmetry of a $2\pi/3$ rotation of the lattice about $\vec{\omega}$, and a $\e^{2 \pi i \sigma_z /3}$ on-site unitary.

The bands of this tight binding model can be topologically non-trivial. We demonstrate this by calculating their Chern numbers within the \emph{local density approximation} (LDA). The LDA neglects the hopping modulation due to Bose enhancement and approximates the tight-binding model as translationally invariant. This approximation remains accurate in a given neighborhood of the lattice provided the modulation is on a length scale much greater than the lattice spacing, i.e. $N\gg 1$. Performing an LDA expansion about the centroid $n_j = N/3$ yields the following tight-binding model on a 2D triangular lattice
\begin{equation}
    H \ket{\vec{n}} = V\ket{\vec{n}} + \tfrac{1}{3} N{\sum}_{j} \left( G_j \ket{\vec{n}+\vec{\delta}_j} + G_j^\dagger \ket{\vec{n}-\vec{\delta}_j}\right)
    \label{eq:bulk_ham}
\end{equation}
where $V = \Delta\sigma_z+\omega N$ and hops are in the directions $\vec{\delta}_j = \vec{e}_{j+1} - \vec{e}_{j-1}$, where $\vec{e}_j$ are the usual Cartesian basis vectors. The corresponding Bloch Hamiltonian may be obtained by Fourier transform of~\eqref{eq:bulk_ham}
\begin{equation}
\label{eq:ham_bloch}
    H(\vec{k}) = \omega N + \tfrac23 g N \vec{\eta}(\vec{k}) \cdot \vec{\sigma}
\end{equation}
where $\vec{\eta}(\vec{k})$ is explicitly given in the SM.

The Bloch Hamiltonian in~\eqref{eq:ham_bloch} corresponds to the QWZ Hamiltonian on a triangular lattice (see SM)~\cite{qi2006topological,bansil2016colloquium}. For certain parameters, the bands of this model are topologically non-trivial, with Chern numbers $C=\pm 1$. In particular, the lower band has $C=-1$ when $3\Delta/2 g N\in[-1,3]$. Thus for any $\Delta$, the lower band is topologically non-trivial for sufficiently large $N$. To maximize the range of $N$ values with non-trivial bulk bands, we set $\Delta=0$.

When the Chern number of the lower band is nonzero, the bulk-edge correspondence promises chiral boundary modes in the bulk energy gap~\cite{haldane1988model,bansil2016colloquium}. These boundary modes are not localized at the geometric edge of the photon lattice due to the inhomogeneous hopping amplitudes. Although the modes have a large weight on the corners of the photon lattice, away from the corners they deviate into the bulk, following the black lines in Fig.~\ref{fig:p1}(b). 

The location of the boundary mode may be calculated within the LDA. We calculate an analogous local band structure to~\eqref{eq:ham_bloch} expanding about any site $\vec{n}$. The chiral boundary mode is localized at the boundary between the topological ($C=-1$) and trivial ($C=0$) regions, where the local band gap closes. The boundary mode position may be calculated, and is given in terms of the dummy variable $x$ by (see SM)
\begin{subequations}
\label{eq:chiral_mode_path}
\begin{equation}
    n_j = N \nu(x - j /3), \quad \nu(x) = \tfrac19 \big(1+2 \cos (2\pi x)\big)^2.
    \end{equation}
This reproduces the shape seen in Fig.~\ref{fig:p1}(b). Moreover, a wavepacket on this boundary circulates, with
\begin{equation}
    x_t = x_0 + t/T, \quad T = \frac{4 \pi}{\sqrt{3} g},
\end{equation}
\end{subequations}
as may be calculated within the LDA (where the wave-packet speed follows from the local Dirac velocity), or from the semi-classical dynamics (see SM).

	\begin{figure}
	
	\includegraphics[width=\columnwidth]{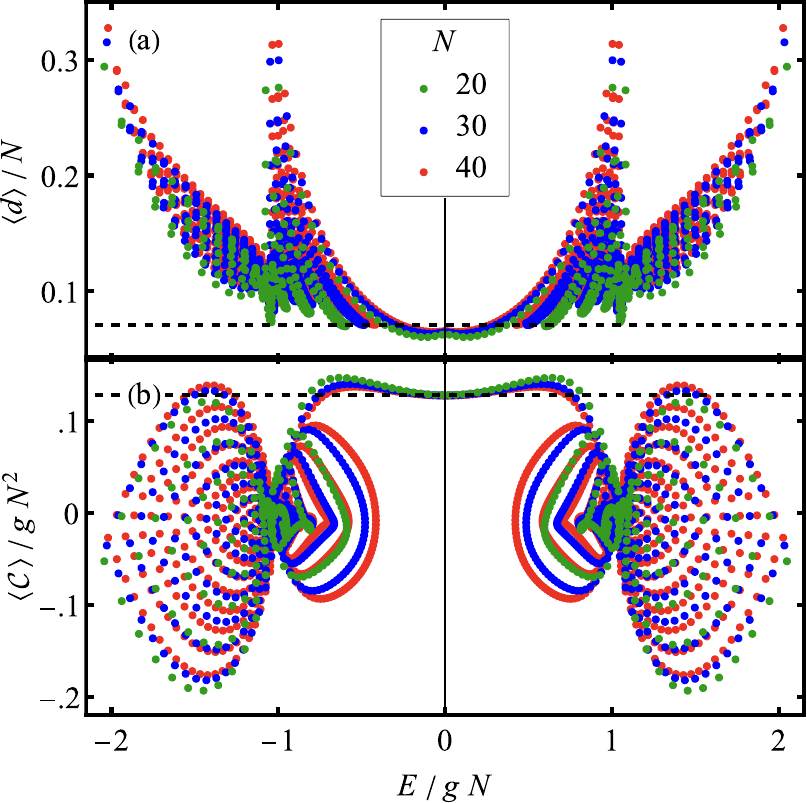}
	
	\caption{\emph{Spectral properties:} (a) Rescaled distance $\cexp{d}$ from the edge of the photon lattice vs energy $E$ for $N=20,30,40$ photons. The band of states crossing $E = 0$ with $\cexp{d}/N$ close to the LDA boundary value (modes) are boundary modes. (b) Rescaled circulation $\cexp{\mathcal{C}}$ versus energy for the same $N$ values. The band of boundary modes is chiral as $\cexp{\mathcal{C}}/gN^2 \neq 0$; indeed the value is close to that predicted by the LDA (dashed). }
	\label{fig:2}
\end{figure}

\paragraph*{Spectral properties:} We compute the spectral properties of $H$ numerically within a number sector $N$, and compare them to the predictions of the LDA. Henceforth, we work in a rotating frame in which $H \to H - \omega N$, so that the middle of the spectrum is at $E= 0$. 

The typical energy scales follow directly from $H$~(Eq.~\eqref{eq:bulk_ham}). Away from the corners of the photon lattice, at least two cavities have $n_j \sim N$ populations. Since $b_j \sim \sqrt{n_j}$, states with low weight at the corners have energies $E \sim g N$. The states located at the corners, where only one cavity has $n_j \sim N$, instead have energies $E \sim g \sqrt{N}$. The spectral gap is thus $O(\sqrt{N})$; numerically $E_\mathrm{gap} \approx 5.3 g\sqrt{N}$ . Crossing this gap is a band of chiral boundary modes, as we show below. 

Define $d_{\vec{n}}$, the Euclidean distance of a site from the nearest geometric edge of the photon lattice in the fixed $N$ plane, given by $d_{\vec{n}} = \sqrt{3/2} \,\min_{j} n_j$. The corresponding operator measuring distance from the edge is then $d = \sum_{\vec{n}\sigma} d_{\vec{n}} \ket{\vec{n}\sigma} \bra{\vec{n}\sigma}$. 

Fig.~\ref{fig:2}(a) plots eigenstate expectation values $\cexp{d}$ vs. their energies for $N=20,30,40$, with both quantities rescaled by $N$. The data collapse of the high energy states with $E \sim N$ is apparent, as is the drift to lower values of $E/gN$ of the lower energy states with $E \sim \sqrt{N}$. The boundary modes appear as a band of states at small constant $\cexp{ d } / N$, approaching the LDA value of $d_\mathrm{bm} /N = 0.071...$ (dashed line in Fig.~\ref{fig:2}(a), see SM). The energy range of the band defines the bulk gap. 

The circulation of probability current on the photon lattice quantifies the chiral transport of photons between cavities. The total photon current into cavity $j$ is measured by $J_j = i[H,\hat{n}_j]$. In a stationary state the expectation value of $\vec{J}$ is zero. Nevertheless, the state may have a non-zero photon current that flows from cavity $j$ to $j+1$ (identifying label $4$ with $1$). On the Fock lattice, these physical photon currents manifest as chiral probability currents. The chirality of the probability currents may be detected by the angular momentum in the direction normal to the Fock plane at constant $N$. We thus define the circulation operator $\mathcal{C}$,
\begin{equation}\label{eq:circ}
    \mathcal{C} = \tfrac{1}{2}\big(\hat{\vec{n}}\times \vec{J}\big) \cdot \vec{u}_\omega+{\rm h.c.}
\end{equation}
where $\vec{u}_\omega$ is the unit vector parallel to $\vec{\omega}$. As $\vec{J}$ and $\vec{n}$ scale with $N$, we expect $\cexp{ \mathcal{C} } \sim N^2$. 

Fig.~\ref{fig:2}(b) plots the eigenstate expectation values of the circulation (re-scaled by $N^2$) versus the rescaled energy. Once again, we see a nearly flat band of states crossing $E=0$ with a large circulation that approaches the LDA predicted value of $\cexp{ \mathcal{C} }/g N^2 = 0.128...$ (see SM). This confirms that the band of boundary modes is chiral. 

\paragraph*{Circulation of Fock and Coherent states:} The boundary modes allow for the chiral transport of quantum states between the cavities. We demonstrate this for two initial states: the first with definite photon number, and the second a superposition across number sectors
\begin{equation}
\ket{\psi_{\mathrm{Fock}}} = \ket{0,0,N}\ket{+}, \quad \ket{\psi_{\mathrm{coh.}}} = \ket{0,0,\alpha}\ket{+}.
\label{eq:initial_states}
\end{equation}
In each case, the qubit is prepared in the $x$-polarized state $\ket{+} \!= \!\tfrac{1}{\sqrt{2}}(\ket{0 } \!+ \!\ket{1})$, cavities 1 and 2 are empty, and cavity 3 is prepared in either a Fock state of $N$ photons, or a coherent state of $|\alpha|^2 = \bar{N}$ photons. 

We first demonstrate that, for sufficiently large $N$, $\ket{\psi_{\mathrm{Fock}}}$ circulates under time evolution \emph{with a period $T$ that is independent of $N$} (up to sub-leading $O(N^{-1})$ corrections, see Fig.~\ref{fig:p1}(c)). 

The circulation of $\ket{\psi_{\mathrm{Fock}}}$ follows from its large overlap with the circulating band. To see this, first note the choice of $\ket{+}$ for the initial qubit state minimizes the energy uncertainty $\Delta E = \sqrt{\cexp{H^2}-\cexp{H}^2}$. By direct calculation one finds, $\ket{\psi_{\mathrm{Fock}}}$ has an average energy $\cexp{H} = 0$ in the center of the bulk gap, and an energy uncertainty $\Delta E/g = (N(4 - 2 \sqrt{3}))^{1/2} = \sqrt{N} \cdot 0.73\ldots$. This uncertainty is much smaller than the bulk gap $E_\mathrm{gap}/g \sim 5.3 \sqrt{N}$. Hence $\ket{\psi_{\mathrm{Fock}}}$ has large overlap with the boundary modes.

Photon circulation following a quench from $\ket{\psi_{\mathrm{Fock}}}$ is shown in Fig.~\ref{fig:p1}(c). In particular, one finds $\langle \hat{n}_j \rangle \approx N$ at times $t/T=m + j/3$, for integer $m$, where $T = 4 \pi/\sqrt{3} g$ is the period (see below). We see that Fock states with $N=20,30,40$ photons circulate with the same $N$-independent period. Intuitively, the $N$-independence follows as the the wavepacket velocity on the photon lattice scales with the current $J \propto N$ due to the Bose enhancement, while the path length of the boundary mode is also $\propto N$. 

The dynamics and period may be calculated semi-classically (see SM) yielding (up to $O(1/N)$ corrections) 
\begin{equation}
\begin{gathered}
    \cexp{\hat{n}_j} \!= \!N \nu \big(t/T \!- \!j/3 \big), \, 
    \cexp{\vec{\sigma}} \!= \!\big( \!\cos( 2 \pi t /T) , \sin (2 \pi t /T),0 \big)
\end{gathered}
\label{eq:semiclassical_dyn}
\end{equation}
with $\nu(\cdot)$ as in~\eqref{eq:chiral_mode_path}. 

 \begin{figure}
\includegraphics[width=\columnwidth]{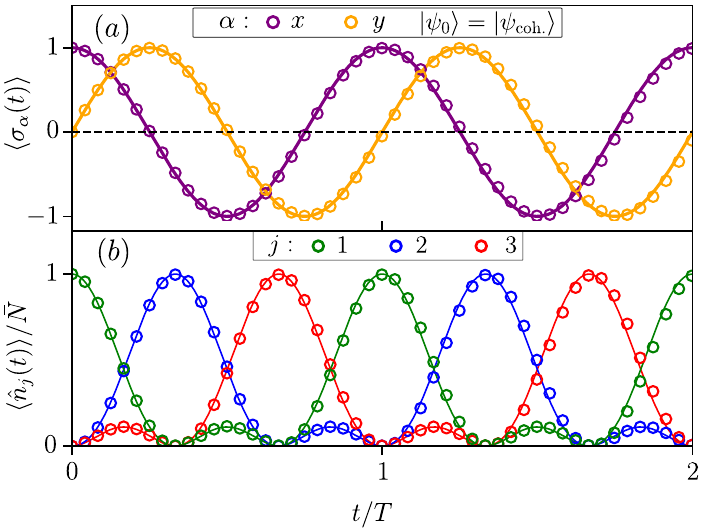}
	
\caption{\emph{Coherent state dynamics:} Dynamics of (a) the qubit and (b) the cavities for an initial coherent state $\ket{\psi_0} = \ket{\psi_{\mathrm{coh.}}}$ with mean photon number $\bar{N} =50$. The solid lines in both panels show the semi-classical predictions~\eqref{eq:semiclassical_dyn}.}
\label{fig:3}
\end{figure}

Numerical simulations confirm the semi-classical predictions. Specifically, Fig.~\ref{fig:3}(a) shows that the qubit state rotates about the $z$-axis with period $T$, in synchrony with the cavity state populations and in perfect agreement with~\eqref{eq:semiclassical_dyn}. Moreover, while the circulation persists, the qubit remains nearly unentangled with the cavities, such that at times $t = nT/3$ the state is approximately rotated by $2\pi n/3$, i.e., $\ket{\psi(n T/3)} \approx U_{\mathrm{C}_3}^n \ket{\psi_{\mathrm{Fock}}}$. 

Semi-classics also suggest that \emph{any coherent superposition of photon number states} in the starting cavity circulates, as the computed period of circulation $T$ is independent of $N$. Fig.~\ref{fig:3}(b) numerically demonstrates this for an initial coherent state $\ket{\psi_{\mathrm{coh.}}}$~\eqref{eq:initial_states}, (cf. Fig.~\ref{fig:p1}(c)).

We note that the circulation does not persist for arbitrarily low photon numbers $N$. At small $N$ the photon lattice contains only a few sites, there is no meaningful bulk or boundary, and the arguments of the previous sections do not hold. In practice, boundary modes with spectral properties as in Fig.~\ref{fig:2} appear for $N \gtrsim 10$. This contrasts with the models in
Refs.~\cite{Roushan:2017aa, Wang:2016aa, cai2021topological} for which, in the absence of perturbation, boundary modes persist to low $N$.

\paragraph*{Lifetime and robustness of circulation:} 

Chiral boundary modes are guaranteed by bulk band topology, and hence persist even upon perturbation. The circulation lifetime is then limited by the dispersion of the boundary wavepacket on the photon lattice. 

For the model~\eqref{eq:ham}, the stroboscopic revivals $\cexp{n_j} \approx N$ decay after a lifetime $\tau \sim TN$. This holds for arbitrary perturbation that preserves $\mathcal{P}$ anti-symmetry. Breaking the $\mathcal{P}$ anti-symmetry shortens the lifetime to $\tau \sim T \sqrt{N}$. To see this, identify the lifetime with the time when the linear size $\Delta n$ of a boundary mode wavepacket spreads across the photon lattice $\Delta n \sim N$. The wavepacket broadens in time due to the non-linear dispersion of the boundary mode, $E_k/N \sim v k + c k^p + \ldots$ where $k$ is the momentum along the boundary. The $\mathcal{P}$ anti-symmetry requires that $E_k$ is odd in $k$, and hence the leading order correction is $p=3$. Without $\mathcal{P}$ anti-symmetry, $p=2$. These corrections to the linear dispersion cause the wavepacket to spread as $\Delta n \sim c N t (\Delta k)^{p-1}$, yielding $\tau \sim (\Delta k)^{1-p}$. Finally, the lifetime is obtained by combining the energy uncertainty of the initial state $\Delta E \sim \sqrt{N}/T$ with the momentum uncertainty of the linear dispersion $\Delta k \sim \Delta E/vN \sim 1/(v T \sqrt{N})$.

\begin{figure}
\includegraphics[width=\columnwidth]{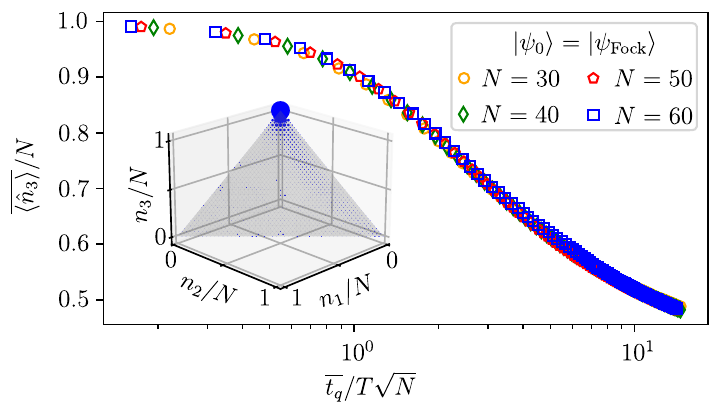}
	
\caption{\emph{Topological robustness:} The sample averaged cavity population revivals $\overline{\braket{\hat{n}_3}}$ at stroboscopic times $\overline{t_q}$ for initial state $\ket{\psi_{\mathrm{Fock}}}$ at different values of $N$. The data collapse confirms that the revivals decay on a timescale $\tau \sim \sqrt{N} T$.
(Inset) The dispersed wavepacket is visualized on the photon lattice at an intermediate stroboscopic time when $\braket{\hat{n}_3}/N\approx 0.9$. The area of each blue dot is proportional to the probability of the wavepacket on the associated photon lattice site.
}
\label{fig:4}
\end{figure}

The lifetime scaling $\tau \sim \sqrt{N} T$ is demonstrated for an initial Fock state~\eqref{eq:initial_states} in Fig.~\ref{fig:4}. Here the $\mathcal{P}$ anti-symmetry is broken by a random perturbation; specifically $G_j \to G_j + \sum_\alpha \delta_{j\alpha}\sigma_\alpha $ for random complex $\delta_{j\alpha}$ with real and imaginary parts drawn independently from the uniform distribution $[-\delta,\delta]$ with $\delta = 0.1$. To obtain Fig.~\ref{fig:4} we numerically calculate the $q$th revival (the max value of $\cexp{\hat{n}_3}$ in the $q$th period) and the corresponding time $t_q$. Finally, we obtain $\overline{\cexp{\hat{n}_3}}$ and $\overline{t_q}$ by averaging over $500$ disorder realizations.
\paragraph*{Floquet-engineered realization:} Three body interactions are generically hard to engineer. The Hamiltonian~\eqref{eq:ham} may however be engineered by applying high frequency drives to a two-body Hamiltonian of a qubit and three cavities. Specifically, ~\eqref{eq:ham} follows from the Hamiltonian below to leading order in the Magnus expansion (see SM):
\begin{equation}
    H_0(t) = \Delta \sigma_z + {\sum}_j \omega_0 b_j^{\dagger}b_j +  {\sum}_j \left( b_{j}^{\dagger} \vec{r}_j(t) \cdot \vec{\sigma} \,  + \mathrm{h.c} \right).
    \label{eq:Floquet_Ham}
\end{equation}
Above, $\vec{r}_j(t) = \vec{r}_{j+} \e^{i \omega_\mathrm{d} t} + \vec{r}_{j-} \e^{- i \omega_\mathrm{d} t}$ define the single tone drive with frequency $\omega_\mathrm{d}$, and $\vec{r}_{j\pm}$ are dimensionless complex constant vectors encoding the drive protocol. As the terms in the Hamiltonian~\eqref{eq:Floquet_Ham} are accessible in current circuit-QED platforms~\cite{cqed_exp1,cqed_exp2,cqed_exp3,exp4}, Floquet engineering may provide an experimentally feasible route to realize the topological circulation of photonic states.

\paragraph{Discussion:}
We presented a cavity-QED device that supports high-fidelity and long-lived quantum state circulation. An arbitrary quantum state (with $N \gg 1$ photons) prepared in a single cavity, circulates between the cavities, periodically recohering in each in turn. This circulation persists for evolution times scaling as $T \sqrt{N}$, where $T$ is the $N$-independent period of the circulation (see Eq.~\eqref{eq:circulator}). The origin of the circulation is topologically protected boundary modes on the photon lattice; the circulation is thus stable to arbitrary small perturbations. 

The model presented is reminiscent of a circulator: a three port linear device in which the input from port $i$ appears in the output port of $i+1 (\operatorname{mod} 3)$. The development of on-chip non reciprocal elements, such as circulators, is an active area of current research~\cite{intro_nonreciprocal2,isolator1,intro_nonreciprocal7,isolator2}. The device studied here circulates power from cavity $i$ to cavity $i+1$ provided the qubit state is prepared in the correct initial state $|\sigma_i\rangle$. Thus this device acts as a circulator only for pulsed input power (with pulse widths $\delta t \ll T$) and provided the qubit is reset between pulses (see SM), and not for an arbitrary continuous input signal.

Cavity-QED models with $m$ cavities map to tight-binding models on the photon lattice with $m$ semi-infinite dimensions. This mapping is a powerful one; when the photon lattice model is characterized by bulk topological invariants, there are interesting associated cavity responses, such as adiabatic~\cite{photon_pumping1,photon_pumping2,photon_pumping3} and non-adiabatic photon pumping~\cite{long2021nonadiabatic}, as well as the cavity state transfer studied here. However, non-trivial topological invariants alone are not sufficient: indeed, the Haldane model on the photon lattice~\cite{cai2021topological,yuan2024quantum} has chiral boundary modes, but these modes do not result in cavity state transfer as they have little overlap with product states of the cavities.

Finally, we note that the photonic circulation described in this work persists in the classical limit (see SM). This raises the intriguing question of whether classical systems can show robust dynamics due to non-trivial topology.

\begin{acknowledgments}
\paragraph{Acknowledgements:} We acknowledge useful discussions with Chris Laumann, Ivar Martin, Mark Rudner, Alicia J Kollár, Pedram Roushan, Dominik Vuina, Michael Kolodrubetz, Daniel Arovas and Maciej Lewenstein. AC thanks the Max Planck Institute for the Physics of Complex Systems for its hospitality. This work was supported by the the National Science Foundation (NSF) under Grant No. DMR-2103658 (SB), and the Air Force Office of Scientific Research (AFOSR) under Grant Nos. FA9550-24-1-0121 (AC), FA9550-20-1-0235 (AC \& SB), AFOSR FA9550-21-1-0342 (SB), and AFOSR Multidisciplinary University Research Initiative (MURI) program Grant No. FA9550-21-1-0069 (PC).
\end{acknowledgments}

%TC:ignore
%\section*{Word Counts}

%\detailtexcount{main}
%TC:endignore

\bibliography{paper-master,bib}
\end{document}

\title{Supplemental material for ``Chiral quantum state circulation from photon lattice topology''}
\author{Souvik Bandyopadhyay}
\affiliation{Department of Physics, Boston University, 590 Commonwealth Avenue, Boston, Massachusetts 02215, USA}

\author{Anushya Chandran}
\affiliation{Department of Physics, Boston University, 590 Commonwealth Avenue, Boston, Massachusetts 02215, USA}

\author{Philip JD Crowley}
\affiliation{Department of Physics, Harvard University, Cambridge, MA 02138, USA}
\affiliation{Department of Physics and Astronomy, Michigan State University, East Lansing, Michigan 48824, USA}
\maketitle

{
  \hypersetup{linkcolor=blue}
  \tableofcontents
}

\section{The tight-binding model on the photon lattice within the local density approximation}\label{sec:1}

In this appendix (i) we derive the LDA Hamiltonian, (ii) we then derive the LDA Hamiltonian in Bloch form in the neighborhood of the centroid $n_j = N/3$  Eq.~(4) of main text, and calculate the associated Chern numbers for the bands of this model, finally (iii) we derive the position of the chiral boundary mode on the photon lattice.

\subsection{Local density approximation (LDA)}

We begin from the Hamiltonian, Eq.~(2),
\begin{equation}
	H = \Delta\sigma_z+\omega\hat{N}+ \sum_{j=1}^3\left(b_{j+1}^{\dagger}b_{j-1}G_{j}+\text{h.c.}\right).
\end{equation}
To make progress we use the phase number representation of the bosonic creation operators
\begin{equation}
    b_j = \sqrt{\hat{n}_j} \e^{i \phi_j}, \qquad [\hat{n}_j, \phi_l] = i \delta_{jl}
\end{equation}
it is straightforward to verify that these satisfy the usual commutation relations $[b_i,b_j^\dagger] = \delta_{ij}$. In the Fock basis these operators act as
\begin{equation}
    \hat{n}_j \ket{\vec{n}} = n_j \ket{\vec{n}}, \qquad \e^{i \phi_j} \ket{\vec{n}} = \ket{\vec{n} - \vec{e}_j}
\end{equation}
where $\vec{e}_j$ are the usual Cartesian basis vectors $\vec{e}_1 = (1,0,0)^T$, $\vec{e}_2 = (0,1,0)^T$, $\vec{e}_3 = (0,0,1)^T$.

For $n_j \gg 1$ the Hamiltonian varies on the length-scale $O(n_j)$, much larger than the photo lattice length. Neglecting this local variation, we may replace the number operators with their local scalar values $\hat{n}_j \to n_j$. This is the local density approximation (LDA)
\begin{equation}
    H(\vec{n}) = \Delta\sigma_z+\omega N + \sum_j \left( f_j(\vec{n}) G_j \e^{ - i \vec{\delta}_j \cdot \vec{\phi}  } + \mathrm{h.c.} \right)
\end{equation}
where, as in the main text, $\vec{\delta}_j = \vec{e}_{j+1} - \vec{e}_{j-1}$, and we have defined $f_j(\vec{n}) = \sqrt{n_{j+1}n_{j-1}}$. Setting $n_j = N/3$, and using that $\e^{- i \vec{\delta}_j \cdot{\vec{\phi}}}\ket{\vec{n}} = \ket{\vec{n}+\vec{\delta}_j}$ one straightforwardly obtains Eq.~(3).

To obtain the Bloch Hamiltonian Eq.~(4) we perform a basis rotation $\vec{k} = O \vec{\phi}$ (for orthogonal $O$) such that $k_x$ and $k_y$ are in the fixed $N$ plane,
\begin{eqnarray}
    k_x = \frac{\phi_1 - \phi_2}{\sqrt{2}}, \qquad k_y = \frac{\phi_1 + \phi_2 - 2 \phi_3}{\sqrt{6}}, \qquad k_z = \frac{\phi_1 + \phi_2 + \phi_3}{\sqrt{3}}
\end{eqnarray}
and a corresponding rotation $\vec{a}_j = O^T\vec{\delta}_j$ to yield
\begin{equation}
    H(\vec{n}) = \Delta\sigma_z+\omega N + \sum_j \left( f_j(\vec{n}) G_j \e^{ - i \vec{k} \cdot \vec{a}_j } + \mathrm{h.c.} \right).
    \label{seq:Hmod}
\end{equation}
in which $k_x, k_y$ and $N$ are good quantum numbers, and $k_z$ does not appear.

\subsection{The LDA Hamiltonian at the centroid $n_j = N/3$ and its Chern number}

The LDA Hamiltonian~\eqref{seq:Hmod} in the vicinity of the centroid $n_j = N/3$ takes the form 
\begin{equation}
    \begin{aligned}
	H &= \Delta\sigma_z+\omega N + \tfrac13 N \sum_j \left(  G_j \e^{ - i \vec{k} \cdot \vec{a}_j } + \mathrm{h.c.} \right)
        \\
        & = \omega N + \tfrac{2}{3}g N\vec{\eta}(\vec{k})\cdot\vec{\sigma},
    \end{aligned}
\end{equation}
with the elements of the Bloch vector $\vec{\eta}$ given by
\begin{equation}
    \begin{aligned}
        \eta_x(\vec{k}) & = \sin (k_x) + \sin (k_x/2)\cos (\sqrt{3}k_y/2)
        \\
        \eta_y(\vec{k}) & = \sqrt{3} \cos (k_x/2)\sin (\sqrt{3}k_y/2)
        \\
        \eta_z(\vec{k}) & = \frac{3\Delta}{2gN}- \cos (k_x) - 2 \cos (k_x/2) \cos(\sqrt{3}k_y/2)
    \end{aligned}
\end{equation}
This representation appears complicated, but has a six-fold rotational symmetry $\vec{\eta}(R_j \vec{k}) = R_j\vec{\eta}( \vec{k})$ where $R_j$ is a rotation by $\pi j/3$ about the $k_z$-axis. 

\begin{figure}[ht]
		\includegraphics[width=\textwidth,height= 6.35cm]{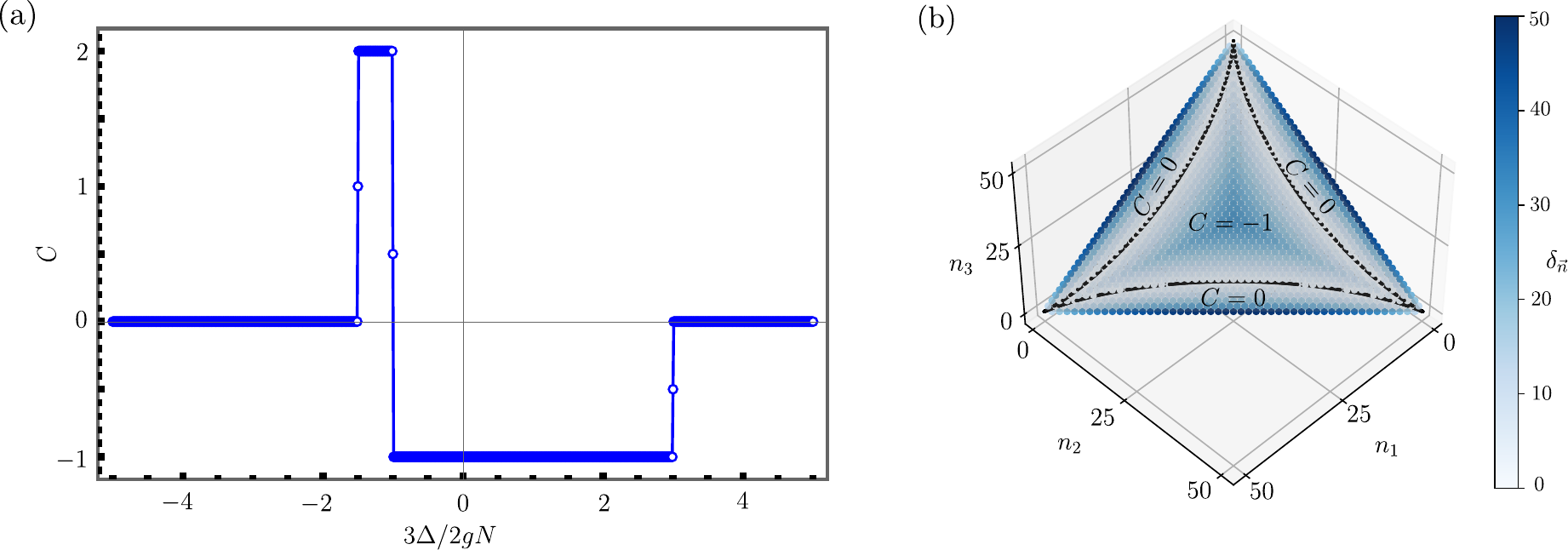}
	\caption{(a) Chern number of the lower band of the isotropic QWZ model vs the renormalized qubit splitting $3\Delta/2gN$. In the large photon number limit $N\gg 1$, $|3\Delta/2gN| \approx 0$ for finite $\Delta$ and the lower band always have Chern number $C=-1$. (b) Local band gap $\delta_{\vec{n}}$ and local Chern number $C$ for $N=50$ sites in Fock space within the LDA. The local band gap is zero on the black curves.}
	\label{fig:sm_phase_diagram}
\end{figure}

The Chern number of the lower band can be calculated using the usual definition 
\begin{equation}
    C=\frac{1}{2\pi}\oint_{BZ} \hat{\eta}(\vec{k}).\left[\partial_{k_x}\hat{\eta}(\vec{k})\times\partial_{k_y}\hat{\eta}(\vec{k})\right]d^2\vec{k},
\end{equation}
where $\hat{\eta}(\vec{k})=\vec{\eta}(\vec{k})/|\vec{\eta}(\vec{k})|$. Fig.~\ref{fig:sm_phase_diagram}(a) plots the Chern number of the lower band vs $3\Delta/2gN$. In the limit $N\to \infty$, at any $\Delta$, the lower band is a Chern band with a Chern number of $-1$. To obtain bulk bands with non-zero Chern numbers at smaller values of $N$, we set $\Delta=0$ in the main text. 
Fig.~\ref{fig:sm_phase_diagram}(b) shows the Chern number of the lower band as a function of $\vec{n}$ on a fixed total photon number surface in the photon lattice. The Chern number transitions from $-1$ in the bulk to $0$ near the lattice edges, across the black curve. The figures also show the local band gap within the LDA in color, which vanishes on the black curve. We observe that the black curves touch the corners of the lattice; for state circulation, it is important that the states $|\psi, 0, 0\rangle$ have overlap with the boundary modes, which are localized near the black curves.

\subsection{The position of the chiral boundary mode $\vec{n}$ on the Photon lattice}

As $\vec{n}$ is varied, one finds that the Chern number of the lower band may change. Specifically, we find that the bands of the LDA Hamiltonian~\eqref{seq:Hmod} are trivial if at least one of the $j$ satisfies
\begin{equation}
    f_j > f_{j+1} + f_{j-1} + \frac{3 \Delta}{2 g N} \quad \implies \quad C = 0
\end{equation}
To verify this, it is straightforward to check that when $f_j = f_{j+1} + f_{j-1} + \frac{3 \Delta}{2 g N}$ the band gap closes due to the appearance of a Dirac point at the $M$-point in the Brillouin zone $\vec{k}_M = (0,2\pi/\sqrt{3})$. 

The position of the chiral boundary mode follows the line of points satisfying the above equality. That is, neglecting the correction from the $\Delta$ term (which is sub-leading in $N$), and recalling $f_j = \sqrt{n_{j+1}n_{j-1}}$, there is a topologically protected chiral boundary mode which follows the equation,
\begin{equation}
    \sqrt{n_{j+1}n_{j-1}} = \sqrt{n_jn_{j+1}} + \sqrt{n_jn_{j-1}}.
    \label{seq:edge}
\end{equation}
This is the black curve shown in Fig~\ref{fig:sm_phase_diagram}. It is straightforward to verify that Eq.~(5) in the main text is the solution to this equation: specifically the condition that~\eqref{seq:edge} is satisfied for at-least one value of $j$ may be re written as
\begin{equation}
    \begin{aligned}
    0 & =\left(\sqrt{n_1n_2} + \sqrt{n_2n_3} + \sqrt{n_3n_1}\right) \prod_{j=1}^3 \left( \sqrt{n_jn_{j+1}} + \sqrt{n_jn_{j-1}} - \sqrt{n_{j+1}n_{j-1}} \right) 
    \\
    & = 2 n_1 n_2 n_3 N - (n_1 n_2)^2 - (n_2 n_3)^2 - (n_3 n_1)^2
    \end{aligned}
\end{equation}
which may be verified by straightforward application of trigonometric identities.

\section{Wavepacket dynamics and classical limit}

In this section, we consider the classical model of the Hamiltonian considered in the main text. The main result of this section is to show that the circulating trajectories exist in the classical limit and to calculate their form. Using this calculation we then reproduce the LDA values of the boundary distance $\cexp{d}$, and boundary mode circulation $\cexp{\mathcal{C}}$.

\subsection{Model and summary of results}

We consider the Hamiltonian $H$ of the classical model in the rotating frame, given by
\begin{equation}
    H = \Delta\sigma_z +\sum_{j=1}^{3} \left( b_{j+1}^*b_{j-1} G_j + \mathrm{h.c} \right),
\end{equation}
where $\vec{\sigma}=(\sigma_x,\sigma_y,\sigma_z)$ and $\vec{b}=(b_1,b_2,b_3)$. These are a classical spin and classical bosons with Poisson brackets
\begin{equation}
    \{b_n,b_m^*\}=-i\delta_{mn}, \qquad \{b_n,b_m\}=0, \qquad \{\sigma_{\alpha},\sigma_{\beta}\}=2\epsilon_{\alpha\beta\gamma}\sigma_{\gamma}
\end{equation}
and as before $G_n=\vec{g}_n \cdot \vec{\sigma}$, where $\vec{g}_n=g(i\cos{(2n\pi/3)},i\sin{(2n\pi/3)},-1)$. The dynamics of this model may be easily understood as the quantum dynamics projected into the product state manifold.

The technical content of the results is two fold: first we show that for $\Delta = \sqrt{3} g/4$ there is a circulating trajectory with initial state and period of circulation (in the large $N$ limit)
\begin{equation}
    \vec{\sigma}_0 = (1,0,0), \qquad \vec{b}_0 = (0,0,\sqrt{N}), \qquad T = T_\infty := \frac{4 \pi }{\sqrt{3} |g|}
\end{equation}
Deviating from this special case, we show that upon detuning $\Delta$ from this special value, the trajectories survive, but are slightly deformed, and correspond to the initial states
\begin{equation}
    \vec{\sigma}_0 = (1,0,0), \qquad \vec{b}_0 = (0,0,\sqrt{N}) + \frac{\epsilon}{16}\sqrt{\frac{3}{N}}
    (1,1,0) + O(\epsilon /N^{3/2})
\end{equation}
where $\epsilon = \frac{4 \Delta}{\sqrt{3}g}-1$ characterizes the deviation from the exact solution. These trajectories circulate with the period
\begin{equation}
    T =  T_\infty \left( 1 - \frac{3\sqrt{3}}{16} \frac{\epsilon}{N} + \frac{9}{512}\frac{\epsilon(6+5 \epsilon)}{N^2} + O(\epsilon/N^3) \right).
\end{equation} 

\subsection{Dynamical equations with $\sigma_z$ = 0}

In general, the equations of motion then follow straightforwardly from Hamilton's equation $\partial_t f = \{f,H\}$ and are given by 
\begin{equation}
    \begin{aligned}
        \partial_t b_j = - i b_{j+1} G_{j-1} - i b_{j-1} G_{j+1}^*, \qquad 
        \partial_t \vec{\sigma} = \vec{\sigma} \times \vec{B}
    \end{aligned}
\end{equation}
where we have defined $\vec{B} = -2 \nabla_{\vec{\sigma}} H$.

To look for circulating trajectories we restrict to $\sigma_z = 0$ which yields some simplifications to the dynamical equations. Specifically, we define the real vector $\Omega_j = \mathrm{Im} \, G_j$ allowing us to recast the cavity dynamics as
\begin{equation}
    \partial_t \vec{b} = \vec{\Omega} \times \vec{b}
\end{equation}
This dynamical equation preserves the phase of $b_j$ and we are at liberty to assume that $b_j = x_j + i p_j$ with $p_j=0$ for all time. A consequence of $b_j$ being real is that the magnetic field is parallel to $\vec{z}$ and of magnitude 
\begin{equation}
    \vec{B} = B_z \vec{z}, \qquad B_z = -2 \Delta - 2 g N + 2 g K, \qquad K = 3 (\vec{u} \cdot \vec{x})^2
\end{equation}
where $\vec{u} = (1,1,1)/\sqrt{3}$ is a fixed unit vector. The time evolution of $\Gamma$ is then given by
\begin{equation}
    \partial_t \Omega_j = i\vec{g}_j \cdot \partial_t \vec{\sigma} = i\vec{g}_j \cdot \vec{B} \times \vec{\sigma} = i B_z \vec{\sigma} \cdot \vec{g}_j \times \vec{z}
\end{equation}
where with some further manipulation it follows that
\begin{equation}
    \partial_t \vec{\Omega} = B_z \, \vec{\Omega} \times \vec{u}, \qquad \vec{\Omega}\cdot\vec{\Omega}= \Omega_0^2 = \tfrac32 g^2, \qquad \vec{\Omega} \cdot \vec{u} = 0.
\end{equation}
In summary we obtain the dynamical equations
\begin{equation}
    \begin{aligned}
        \partial_t \vec{x} & = \vec{\Omega} \times \vec{x}
        \\
        \partial_t \vec{\Omega} & = B_z \vec{\Omega} \times \vec{u}
        \\
        B_z & = -2 \Delta - 2 g N + 2 g K
        \\
        K & = ( \vec{u} \cdot \vec{x} )^2
    \end{aligned}
    \label{seq:dyn}
\end{equation} 
\subsection{Solutions to the dynamical equations for $\sigma_z = 0$, Eq.~\eqref{seq:dyn}}

The equations~\eqref{seq:dyn} can be simplified by transforming to the frame co-rotating with $\Omega$. Defining primed coordinates $ \vec{x}' = R \vec{x}$, $\vec{\Omega}' = R \vec{\Omega}$ etc, we obtain
\begin{equation}
    \begin{aligned}
        \partial_t \vec{x}' & = ( B_z \vec{u} + \vec{\Omega}') \times \vec{x}'
        \\
        \partial_t \vec{\Omega}' & = 0
        \\
        B_z & = -2 \Delta - 2 g N + 2 g K 
        \\
        K & = 3 ( \vec{u} \cdot \vec{x}' )^2 
    \end{aligned}
\end{equation}
As $\vec{\Omega}'$ is static, we have $\vec{\Omega}' = \vec{\Omega}_0$ for all time. These equations of motion have a static solution $\partial_t \vec{x}' = 0$ (i.e. in which $\vec{x}$ is co-rotating with $\vec\Omega$) when $\vec{x}' \propto B_z \vec{u} + \vec{\Omega}_0$. As the normalization of the initial state is fixed $|\vec{x}'|^2 = N$, this yields
\begin{equation}
    \vec{x}' = \vec{x}_0 = - \operatorname{sign}(g)  \sqrt{N} \frac{B_z \vec{u} + \vec{\Omega}_0}{\sqrt{B_z^2 + \Omega_0^2}}
    \label{eq:xprime}
\end{equation}
Here the sign has been chosen to yield $\vec{x}_0 \to (0,0,\sqrt{N})$ in the limit of large $N$. For this state we have
\begin{equation}
    K = \frac{3 N B_z^2}{B_z^2 + \Omega_0^2}
\end{equation}
and hence the self consistency condition
\begin{equation}
    B_z = -2 \Delta - 2 g N + 6 g N \frac{B_z^2}{B_z^2 + \Omega_0^2} 
\end{equation}
The roots to this equation can be calculated. For $\Delta = - \tfrac12 \Omega_0$ (where we fix the sign convention $\Omega_0 = - \sqrt{3} g/2$) there is an exact root $B_z = \Omega_0$. Expanding about the neighborhood of this root $\Delta = - \tfrac12 \Omega_0 (1 + \epsilon)$ in powers of $N$ we obtain the series expansion for the root
\begin{equation}
    B_z = \Omega_0 \left( 1 + \frac{3 \sqrt{3}}{16} \frac{\epsilon}{N} - \frac{27}{256} \frac{\epsilon(1-\epsilon/6)}{N^2} + O(\epsilon/N^3) \right), \qquad \text{where} \qquad \epsilon = \frac{4 \Delta}{\sqrt{3} g} -1.
\end{equation}
The period of the circulation is then given by
\begin{equation}
    T = \frac{2 \pi}{|B_z|} = \frac{4 \pi }{\sqrt{3} |g|} \left( 1 - \frac{3\sqrt{3}}{16} \cdot \frac{\epsilon}{N} + \frac{9}{512}\frac{\epsilon(6+5 \epsilon)}{N^2} + O(\epsilon/N^3) \right),
\end{equation} 
which indicates the presence of corrections to the circulation frequency. Substituting the expression for $B_z$ into Eq.~\eqref{eq:xprime} yields
\begin{equation}
    \vec{x}_0 = (0,0,\sqrt{N}) + \frac{\epsilon}{16}\sqrt{\frac{3}{N}}
    (1,1,0) + O(\epsilon /N^{3/2})
\end{equation}
which is our final result.

\subsection{Boundary distance $\cexp{d}$ and circulation $\cexp{\mathcal{C}}$}

Using the semi-classical trajectories, we are able to recover the LDA values of the distance of the boundary mode from the edge
\begin{equation}
    \begin{aligned}
        \cexp{d} & = \frac{1}{T} \int_0^T dt  \, d(\vec{n}(t)) 
    \end{aligned}
\end{equation}
where $d(\vec{n}) = \sqrt{3/2} \,\min_{j} n_j$ as in the main text. This integral may be directly evaluated for the semi-classical trajectories (given by Eq. 5) and one obtains
\begin{equation}
    \begin{aligned}
        \cexp{d} & = \frac{1}{T/3} \int_{T/3}^{2T/3} N \nu(t / T) d t 
        \\
        & = N \left( \frac{1}{\sqrt{6}} - \frac{3}{2\sqrt{2} \pi} \right) 
        \\
        & = N \times 0.0707\ldots
    \end{aligned}
\end{equation}
similarly the trajectory averaged circulation is given by
\begin{equation}
    \begin{aligned}
    \cexp{\mathcal{C}} & = \frac{1}{T} \int_0^T dt \, (\dot{\vec{n}} \times \vec{n}) \cdot \vec{u}_\omega
    \\
    & = \frac{1}{T} \int_0^T dt \frac{16\pi N^2}{27 T}\sin^2 \left( \frac{3 \pi t}{T}\right)
    \\
    & = \frac{8 \pi N^2}{27 T}
    \\
    & = \frac{2 g N^2}{9 \sqrt{3}}
    \\
    & = g N^2 \times 0.1283 \ldots
    \end{aligned}
\end{equation}
as quoted in the main text.

\section{Lifetime and robustness of circulation}

In this section, we explain the lifetime of circulation starting from a Fock state and the effect of random photon number conserving perturbations that (i) breaks the $\mathcal{P}$-symmetry and (ii) preserves the $\mathcal{P}$-symmetry.  For (i), we show that the lifetime of circulation scales as $\sqrt{N}$ with the total photon number and for (ii), we obtain even longer lifetimes scaling linearly with $N$.\\

\subsection{Circulation lifetime with $\mathcal{P}$-symmetric perturbations}

We begin by considering the boundary mode in the presence of $\mathcal{P}$-symmetric perturbations. Maintaining the $\mathcal{P}$-symmetry entails that the energy dispersion must be odd in the quasi-momentum $k$ along the boundary, while we know  that the energy scale must be  $O(N)$. Thus we obtain a leading order expansion about $k=0$ given by
\begin{equation}\label{eq:sym_dis_1}
    E(k) = N\left[vk+ck^3+O(k^5)\right],
\end{equation}
where the cubic correction to linear dispersion results in the broadening of the wavepacket moving along the edge.

Circulating states of the form considered in the main text, such as $\ket{\psi_{\mathrm{Fock}}}$, have an energy width $\Delta E\propto \sqrt{N}$ centered at $E=0$, and hence by~\eqref{eq:sym_dis_1} a momentum width $\Delta k\sim 1/\sqrt{N}$ centered at $k=0$. 

The real space dynamics of a wavepacket evolving under this dispersion is captured by considering the Heisenberg evolution of the position operator
\begin{equation}
    x(t) = \e^{- i E(k) t} x_0 \e^{ i E(k) t}  =  x_0 + t E'(k) \approx x_0 + N v t + 3 c t N k^2 + O( t k^4)
\end{equation}
from which the width $\Delta x = \sqrt{\cexp{x^2} - \cexp{x}^2}$ of a wavepacket follows
\begin{equation}
    \begin{aligned}
        (\Delta x)^2 & = (\Delta x_0)^2 + (3 N c t \Delta(k^2))^2 + \ldots.
    \end{aligned}
\end{equation}
This is further simplified by noting that the characteristic scale of $\Delta k^2$ is given by $\Delta (k^2) \sim (\Delta k)^2$.

We now consider the growth of $\Delta x$ for the specific initial wavepackets of the form considered in the main text. Specifically, we consider an initial state $\ket{\psi_{\mathrm{Fock}}}$. The width of this wavepacket thus grows as 
\begin{equation}
    \Delta x \sim N t (\Delta k)^2 \sim t.
\end{equation}
where we neglected $O(1)$ constants. This implies that the rate of dispersion is independent of $N$. 

This result may be summarized in terms of the circulation lifetime. We define the lifetime of circulation $t^*$ as the time when the initial wavepacket spreads extensively on the edge, i.e., $\Delta x(t^*)\sim N$. Thus we have shown that for the $\mathcal{P}-$symmetric model, in the limit of large $N$, the circulation lifetime grows linearly with $N$
\begin{equation}
    t^* \sim N.
\end{equation}

\subsection{Circulation lifetime with $\mathcal{P}$-symmetry breaking perturbations}

In the presence of $\mathcal{P}-$symmetry breaking perturbations the analysis proceeds along the same lines as in the previous section, with the important difference that the quadratic correction to the dispersion is now symmetry allowed. Specifically, we have a dispersion
\begin{equation}\label{eq:sym_dis}
    E(k) = N\left[vk+bk^2+O(k^3)\right].
\end{equation}
and a wavepacket width
\begin{equation}
    (\Delta x)^2 = (\Delta x_0)^2 + (2 n b t \Delta k)^2 + \ldots
\end{equation}
and hence the wavepacket broadens at a faster rate,
\begin{equation}
    \Delta x(t) \sim 2Nbt\Delta k\sim 2\sqrt{N}bt
\end{equation}
This leads to a shorter lifetime of circulation $t^*$ given by
\begin{equation}
    t^* \sim \sqrt{N}.
\end{equation}

\subsection{Numerical verification of circulation lifetimes}

In this section we numerically verify the robustness of circulation. Specifically, we verify its lifetime in both the unperturbed model, and in the presence of three different types of perturbations:  (i) perturbation to the natural frequencies of the cavities that breaks $\mathcal{P}$-symmetry, (ii) perturbation to the cavity-qubit interactions that breaks $\mathcal{P}$-symmetry and (iii) perturbation to the cavity-qubit interactions that respects $\mathcal{P}$-symmetry. All three of these perturbations break the $C_3$ symmetry of the unperturbed Hamiltonian but conserve the total number of photons $N$.\\

\begin{figure*}
		\includegraphics[width=0.39\textwidth]{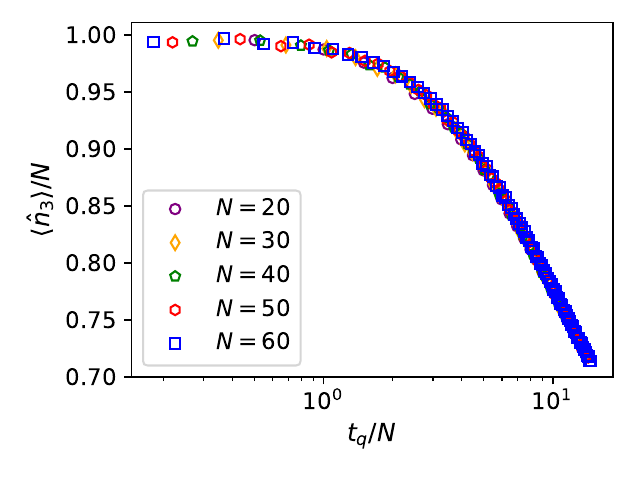}
			
	\caption{Population of cavity $3$ at stroboscopic instants $t_q$, starting from the initial state $\ket{\psi_{\mathrm{Fock}}}$ in the unperturbed model. The scaling collapse for different photon numbers $N$, indicate that the lifetime of circulation scale as $t^*\sim N$.}
	\label{fig:clean_lifetime}
\end{figure*}

\paragraph*{\textbf{Unperturbed model:}} In the unperturbed model we expect the $\mathcal{P}$-symmetry to lead to circulation with a lifetime $t^*\sim N$. In Fig.~\ref{fig:clean_lifetime} we plot the cavity population $\braket{\hat{n}_3}$ at stroboscopic times $t_q=qT$, where $q\in\mathbb{Z}^+$ and $T$ is the time period of circulation. The collapse of $\braket{\hat{n}_3}/N$ vs $t_q/N$ for different $N$, verifies the expected lifetime $t^*\sim N$.\\

\begin{figure}
		\includegraphics[width=\textwidth]{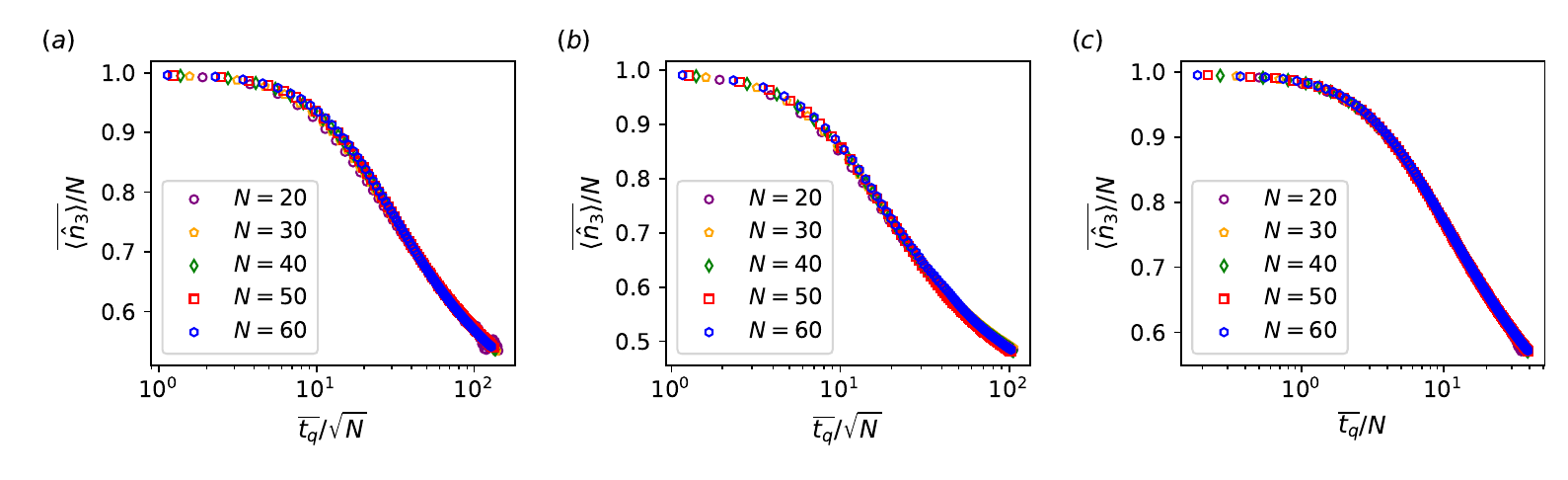}
			
	\caption{Scaled stroboscopic population of cavity-$3$ vs average stroboscopic time $\overline{t_q}$ with, (a) random perturbations in cavity frequencies breaking $\mathcal{P}-$symmetry, (b)  random perturbations in the cavity-qubit couplings breaking $\mathcal{P}-$symmetry and (c) $\mathcal{P}-$symmetric random perturbations in the cavity-qubit coupling. From the scaling collapse in each panel, we extract $t^*\propto\sqrt{N}$, $t^*\propto\sqrt{N}$ and $t^*\propto N$ for (a), (b) and (c) respectively. The initial state for all simulations were set to $\ket{\psi_{\mathrm{Fock}}}=\ket{0,0,N}\ket{+}$. All data is averaged over $500$ realizations with perturbations sampled randomly from a uniform distribution in $[-\delta,\delta]$, such that $\delta=0.25,0.1,0.1$ for (a), (b) and (c), respectively.}
	\label{fig:noisy_lifetime}
\end{figure}

\paragraph*{\textbf{Random perturbations in cavity frequencies:}} Next, we consider perturbations to the cavity frequencies. For unequal cavity frequencies, the effective electric field in Fock space in the direction $(\omega_1,\omega_2,\omega_3)$ is not perpendicular to the plane of motion $n_1+n_2+n_3=N$. This results in an in-plane component of the effective electric field in Fock space which clearly breaks the permutation symmetry between the cavities.

Specifically we consider a perturbation to the Hamiltonian $H \to H + \Delta H$ of the form
\begin{equation}
    \Delta H = \vec{\delta \omega} \cdot \vec{n}
\end{equation}
where the components of the vector $\vec{\delta \omega}$ are sampled independently from the uniform distribution $[-\delta,\delta]$. 

In Fig.~\ref{fig:noisy_lifetime}(a), we show the numerically calculated stroboscopic occupation number in cavity $3$ given an initial state $\ket{\psi_\mathrm{Fock}} = \ket{0,0,N}\ket{+}$. Each series is averaged over random realizations. For each trajectory we numerically calculate the stroboscopic times $t_q$ as the times for which the cavity population $\braket{\hat{n}_3}$ peaks. The corresponding values $t_q$, and $\braket{\hat{n}_3(t_q)}$ are then averaged over disorder realizations to obtain $\overline{t_q}$, and $\overline{\braket{\hat{n}_3}}$. As this perturbation breaks the $\mathcal{P}-$symmetry, we expect a circulation lifetime $t^* \propto\sqrt{N}$. This is verified in Fig.~\ref{fig:noisy_lifetime}(a). \\

\paragraph*{\textbf{Generic random perturbations in qubit-cavity coupling:}} Next, we add a random perturbation to the cavity-qubit coupling that also breaks the $\mathcal{P}-$symmetry of the model. Consider the perturbed Hamiltonian,
\begin{equation}
    \Delta H = \sum_j \left( b_j^{\dagger}b_{j+1} \delta G_{j-1}  + h.c \right), \qquad \delta G_{j} = \vec{\delta g}_j \cdot \sigma
    \label{seq:cavity_perturbs}
\end{equation}
where each of the elements of each vector $\vec{\delta g}_j$ are independent and identically distributed random complex numbers with real and imaginary parts drawn from the distribution $[-\delta,\delta]$. As this perturbation breaks the $\mathcal{P}-$symmetry, we expect a circulation lifetime $t^* \propto\sqrt{N}$. This is verified in Fig~\ref{fig:noisy_lifetime}(b). The numerical averaging procedure is as in the previous case. \\

\paragraph*{\textbf{$\mathcal{P}-$symmetric random perturbations in qubit-cavity coupling:}} Finally, we consider a random perturbation to the cavity-qubit coupling that preserves the $\mathcal{P}-$symmetry of the model. Specifically, we consider the same perturbation as previously~\eqref{seq:cavity_perturbs} with the additional condition $\mathrm{Re} \, \delta g_{jx} = \mathrm{Re}\, \delta g_{jy} = \mathrm{Im} \, \delta g_{jz} = 0$ (whereas, as before $\mathrm{Im} \, \delta g_{jx} , \mathrm{Im}\, \delta g_{jy},  \mathrm{Re} \, \delta g_{jz}$ are independent and identically distributed random numbers drawn from the distribution $[-\delta,\delta]$). As this perturbation preserves the $\mathcal{P}-$symmetry, we expect a circulation lifetime $t^* \propto N$. This is verified in Fig.~\ref{fig:noisy_lifetime}(c). The numerical averaging procedure is as in the previous cases.

\section{Floquet generation from 2-body interactions} 

In this section, we show that the three-body terms in the Hamiltonian (Eq.~(2) of main text) may be Floquet engineered from a bare Hamiltonian involving only two-body interaction terms. This provides a route to experimentally realize the Hamiltonian. 

The results of this section are obtained in two parts: first, the Floquet Hamiltonian is constructed via a leading order high frequency expansion and shown to take the desired form; second, we give the conditions for the validity of this leading order expansion.

\paragraph*{\textbf{Floquet Hamiltonian:}} Consider the periodically driven Hamiltonian with time-period $T_d=2\pi/\omega_d$,

\begin{equation}
    H_{\mathrm{lab}}(t) =  \Delta_0\sigma_z + \sum\limits_{j=1}^3 \omega_0 b_{j}^{\dagger}b_{j}+ \left[\left(\vec{A}_{j}\cos{\omega_dt}+\vec{B}_{j}\sin{\omega_dt}\right)\cdot \vec{\sigma}\,b_{j}^{\dagger}+\mathrm{h.c.}\right],
    \label{eq:high_freq}
\end{equation}
where $\vec{A}_{j}=(A^x_{j},A^y_{j},A^z_{j})$ and $\vec{B}_{j}=(B^x_{j},B^y_{j},B^z_{j})$ are complex-valued vectors. As the couplings $G_{j}$ in the required Hamiltonian are constrained by a $3$-fold rotational symmetry, we require that,
\begin{equation}\label{sec:sm_symm}
        R_{z}(2\pi/3)\vec{A}_{j} = \vec{A}_{j+1},~R_{z}(2\pi/3)\vec{B}_{j} = \vec{B}_{j+1}
\end{equation}
where,
\begin{equation}
    R_z(\theta) = \begin{pmatrix}
\cos{\theta} & -\sin{\theta} & 0\\
\sin{\theta} & \cos{\theta} & 0\\
0 & 0 & 1
\end{pmatrix}
\end{equation}
rotates a Cartesian vector in 3d about the $z-$axis by the angle $\theta$.
To simplify calculations, we introduce the new coordinates,
\begin{equation}
    \begin{split}
    \vec{r}_{j,+} = \frac{\vec{A}_{j}+i\vec{B}_{j}}{2},\\
    \vec{r}_{j,-} = \frac{\vec{A}_{j}-i\vec{B}_{j}}{2}
    \end{split}
\end{equation}
We compute the stroboscopic action of the Hamiltonian $H_{\mathrm{lab}}(t)$ under the high-frequency expansion. To first order in the inverse driving frequency $\omega_d$, we obtain the Floquet Hamiltonian
\begin{equation}\label{eq: mag}
    \begin{aligned}
        H^F & \approx H_{(0)} + H_{(1)} + H_{(2)} + \ldots
        \\ 
        & = H_{(0)} + \frac{1}{\omega_d}[ H_- , H_+] + \frac{5}{6\omega_d^2} \left( [H_+ , [H_0,H_-]] + [H_- , [H_0,H_+]] \right) + \cdots
    \end{aligned}
\end{equation}
where,
\begin{equation}
	H_{\pm} = \frac{1}{T_d}\int_0^{T_d}H_{\mathrm{lab}}(t)e^{\pm i\omega_d t}dt = \sum\limits_{j=1}^3 \left[ b_{j}^{\dagger}(\vec{r}_{j\pm}\cdot \vec{\sigma})+\mathrm{h.c.} \right]
\end{equation}
and $H_{(0)}$ is the average Hamiltonian,
\begin{equation}
    H_{(0)} = \frac{1}{T_d}\int_0^{T_d} H_{\mathrm{lab}}(t)dt = \Delta_0\sigma_z + \sum\limits_{j=1}^3 \omega_0 b_{j}^{\dagger}b_{j}.
\end{equation}

We now go to a rotating frame of reference generated by the unitary $U=\exp\left[i\omega_0\sum_jb_j^{\dagger}b_j\right]$. If $\omega_0$ is the largest energy scale in $H^F$, we can ignore fast oscillating particle number non-conserving terms like $b_{i}^{\dagger}b_{j}^{\dagger}e^{2i\omega_0 t}+\mathrm{h.c.}$ and $b_{j}e^{-i\omega_0 t}+\mathrm{h.c.}$ in the rotating frame. This leads to the effective three-body Floquet Hamiltonian which conserves particle number as,
\begin{equation}\label{eq:effective}
    \tilde{H}^F \approx H_{(0)} + \sum_{ij=1}^{3}\left(\vec{\alpha}_{ij}\cdot \vec{\sigma}b^{\dagger}_{i}b_{j}+\mathrm{h.c.}\right)+\vec{h}\cdot \vec{\sigma}+\mathrm{cons.}
\end{equation} 
 where $\vec{\alpha}_{ij}$ and $\vec{h}$ depend on the harmonics $\vec{r}_{j\pm}$ as,
\begin{equation}
        \vec{\alpha}_{ij} = \frac{2i}{\omega_\mathrm{d}}\left[\vec{r}_{i,-}\times(\vec{r}_{j,-})^*-\vec{r}_{i,+}\times(\vec{r}_{j,+})^*\right],
        \qquad
        \vec{h}  = \tfrac12 \sum_{j=1}^3 \alpha_{jj}
\end{equation} 
In order to generate the target Hamiltonian (Eq.~(2) of the main text), these harmonics need to satisfy
 \begin{equation}\label{eq:floquet_eqns}
 	\vec{\alpha}_{ij} \cdot \sigma  = \delta_{j,i+1}G_{i-1} + \delta_{j,i-1}G_{i+1}^\dagger + \delta_{ij} \delta\Delta \sigma_z
 \end{equation}
where the value $\delta\Delta$ can be absorbed into the value of the bare qubit splitting. Moreover, one can show that for all solutions to these equations we have $\delta\Delta = 2g$ so that we obtain the desired Floquet Hamiltonian
\begin{equation}\label{eq:sm_floquet}
    \begin{aligned}
        \tilde{H}^F & \approx \Delta_0 \sigma_z + \sum\limits_{j=1}^3 \omega_0 b_{j}^{\dagger}b_{j} + \sum_{j}\left[\left(b_j^{\dagger}b_{j+1}G_{j-1} + \mathrm{h.c.}\right) + 2gb_j^{\dagger}b_j\sigma_z \right] + 3g\sigma_z
        \\
        & = \Delta \sigma_z + \omega_0 N + \sum_{j}\left(b_j^{\dagger}b_{j+1}G_{j-1} + \mathrm{h.c.}\right) 
    \end{aligned}
\end{equation}
where $\Delta  = \Delta_0 + g (2 N + 3)$; thus to obtain the Hamiltonian in the main text with $\Delta = 0$ we set $\Delta_0 = - g (2 N +3)$.

Finally we discuss values of $\vec{A}_j$ and $\vec{B}_j$ that are solutions to~\eqref{eq:floquet_eqns}, and hence yield~\eqref{eq:sm_floquet}. There are 12 independent variables (the real and imaginary parts of $\vec{A}_3$ and $\vec{B}_3$), for a solution to~\eqref{eq:floquet_eqns} we need to fix $6$ of them. This counting can be seen by noting that (i) $R_z(2\pi/3)\vec{\alpha}_{ij} = \vec{\alpha}_{i+1,j+1}$, (ii) $\vec{\alpha}_{ij}^* = \vec{\alpha}_{ji}$, and (iii) $\vec{\alpha}_{jj} = - 2 \mathrm{Re}\,\vec{\alpha}_{j+1,j-1}$. Consequently all elements of $\vec{\alpha}_{ij}$ are linear combinations of the 6 degrees of freedom comprising $\vec{\alpha}_{12}$, and we expect a 6-dimensional manifold of solutions. One such solution is given explicitly by
\begin{equation}
    \vec{A}_3 = \operatorname{sign}(g\omega_d) \sqrt{\frac{|g \omega_\mathrm{d}|}{6}}\begin{pmatrix}
        \sqrt{3} \\ \sqrt{3} \\ -i
        \end{pmatrix},
        \qquad 
    \vec{B}_3 = \sqrt{\frac{|g \omega_\mathrm{d}|}{6}}  \begin{pmatrix}
        \sqrt{3} \\ -\sqrt{3} \\ i
    \end{pmatrix}
\end{equation}
where other $A_j/B_j$ are obtained by rotating about the $z$-axis as per~\eqref{sec:sm_symm}.\\

\begin{figure*}
    \centering
\subfigure[]{
    \includegraphics[width=0.3\textwidth, height=6.5cm]{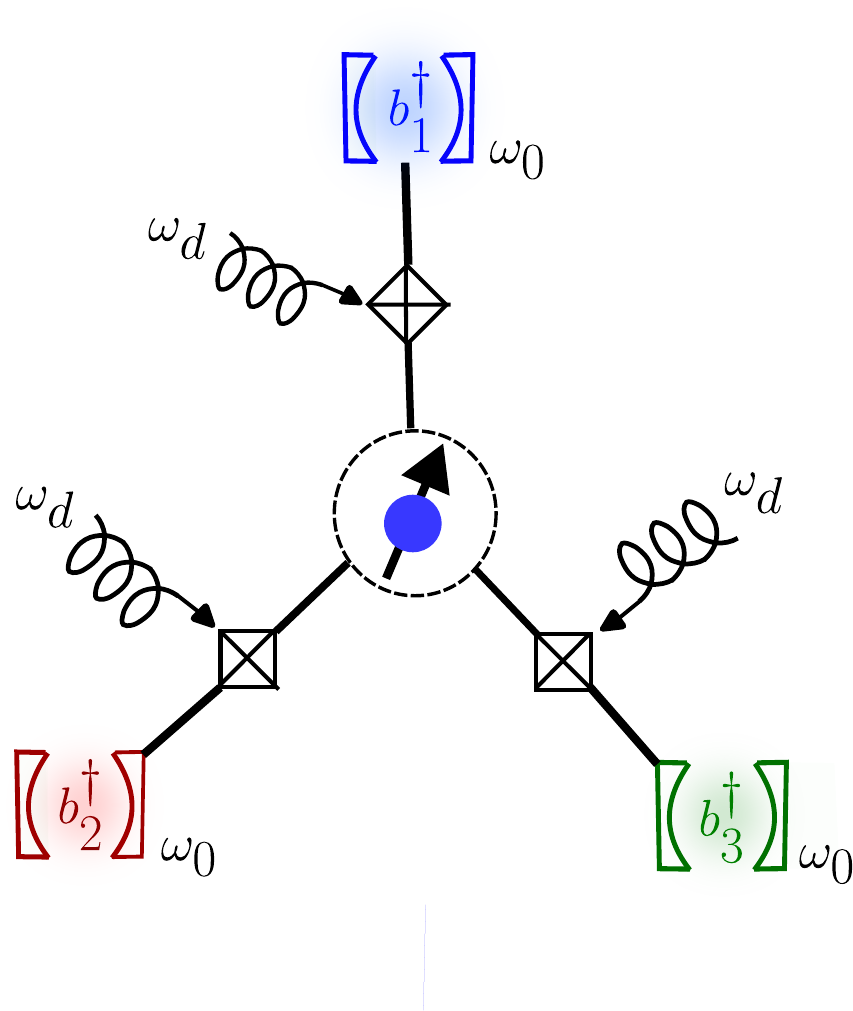}
    }	
    \hspace{0.75cm}
    \subfigure[]{
    \includegraphics[width=0.45\textwidth, height=6.25cm]{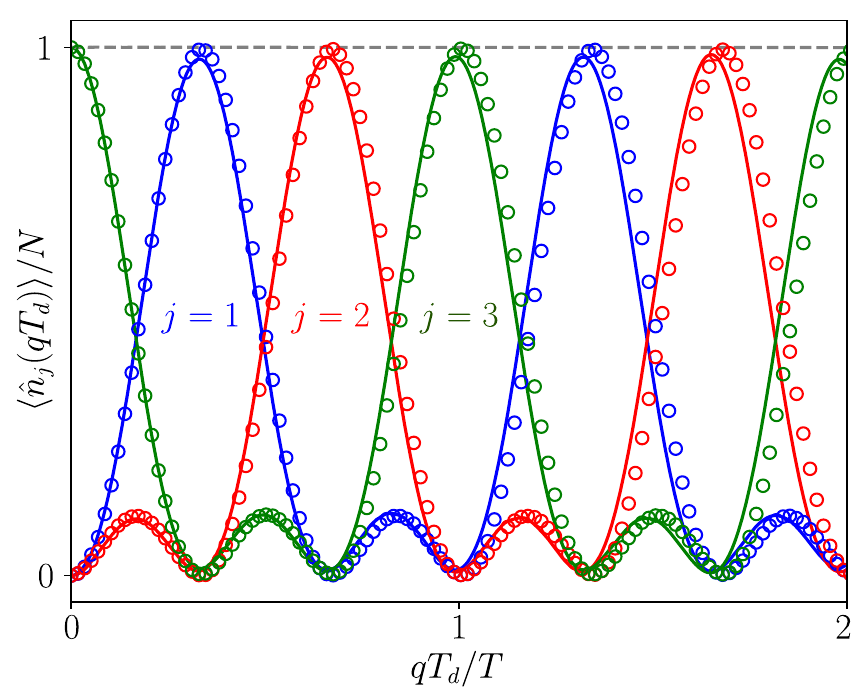}
    }
	\caption{(a) A schematic of the periodic drive. (b) Stroboscopic dynamics at times $t=qT_d$ where $q\in\mathbb{Z}^+$, in the lab frame for the time-periodic Hamiltonian with $2-$qubit interactions (see Eq.~\eqref{eq:high_freq}). The simulation is performed for a natural frequency $\omega_0=10$, driving frequency $\omega_d=5\times 10^3$ and $\Delta_0=-g(2N+3)$, starting from the initial state $\ket{0,0,N}\ket{+}$. The discrete dots indicate dynamics under the static Hamiltonian $H$ (Eq.~(2) of main text) for $\Delta=0$. $T=4\pi/\sqrt{3}g$ is the circulation period and $T_d=2\pi/\omega_d$ is the time period of the drive. The horizontal dashed line shows the total occupation number rescaled by $N$.   
    }
	\label{fig:floquet}
\end{figure*}

\paragraph*{\textbf{Condition for the validity of leading order high frequency expansion and rotating wave approximation:}}
For convergence of the high-frequency expansion, first we impose $||H_{(2)}||\ll||H_{(1)}||$, where $\norm{x} = \sqrt{{\rm tr}[x^{\dagger}x]}$ is the Hilbert-Schmidt norm. Focusing on the particle-number conserving terms in $H_{(1)}$, the norm of the first order correction to the average Hamiltonian for $N\gg 1$ is,
\begin{equation}
    ||H_{(1)}||
    = 
    \norm{ \sum_{j}\left[\left(b_j^{\dagger}b_{j+1}G_{j-1} + \mathrm{h.c.}\right) + g(2N+3)\sigma_z
    \right]
    }
    \sim 
    g\sqrt{N^2+(2N+3)^2 + 2N(2N+3)}
    \sim
    3gN,
\end{equation}
where we have used that energy eigenvalues of the Hamiltonian $H$ defined as Eq.~(2) of main text scales as $E\sim gN$.\\

The second-order correction ($||H_{(2)}||$) is proportional to the commutators $(5/6\omega_d^2)([H_{+},[H_{(0)},H_{-}]]+[H_{-},[H_{(0)},H_{+}]])$, which contain terms of the order $O(g\Delta_0/\omega_d)$ and $O(gN\Delta_0/\omega_d)$. The number conserving terms in $H_{(2)}/N$ in leading order for $N\gg1$ are,
\begin{equation}
    \begin{aligned}
        \frac{H_{(2)}}{N} \approx & \frac{5}{6\omega_d^2N}\sum\limits_{jl}\left(\Delta_0[\vec{r}_{l,+}\cdot \vec{\sigma},(\vec{r}_{j,+})^*.[\sigma_z,\vec{\sigma}]]-\omega_0[\vec{r}_{l,+}\cdot \vec{\sigma},(\vec{r}_{j,+})^*\cdot \vec{\sigma}]\right)b_{l}^{\dagger}b_{j}
        \\
        & \,\, 
        +\frac{5}{6\omega_d^2N}\sum\limits_{jl}\left[ \left(\Delta_0[(\vec{r}_{l,-})^*\cdot \vec{\sigma},\vec{r}_{j,-}.[\sigma_z,\vec{\sigma}]]+\omega_0[(\vec{r}_{l,-})^*\cdot \vec{\sigma},\vec{r}_{j,-}\cdot \vec{\sigma}]\right)b_{l}b_{j}^{\dagger} + \mathrm{h.c. } \right] +O\left(\frac{g\Delta_0}{N\omega_d}\right)
    \end{aligned}
\end{equation}
which can be regrouped as,
\begin{equation}
    \begin{aligned}
        \frac{H_{(2)}}{N} \approx & \frac{10i\Delta_0}{3\omega_d^2N}\sum\limits_{jl}\{\vec{r}_{l,+}\times((\vec{r}_{j,+})^*\times\hat{z})\}\cdot \vec{\sigma}b_{l}^{\dagger}b_{j}+\{(\vec{r}_{l,-})^*\times(\vec{r}_{j,-}\times\hat{z})\}\cdot \vec{\sigma}b_{l}b_{j}^{\dagger}
        \\
        & \,\,
        +\frac{5i\omega_0}{3\omega_d^2N}\sum\limits_{jl}\left[\{(\vec{r}_{l,-})^*\times\vec{r}_{j,-}\}\cdot \vec{\sigma}b_{l}b_{j}^{\dagger}+\{(\vec{r}_{j,+})^*\times\vec{r}_{l,+}\}\cdot \vec{\sigma}b_{l}^{\dagger}b_{j} + \mathrm{h.c.} \right] +O\left(\frac{g\Delta_0}{N\omega_d}\right),
    \end{aligned}
\end{equation}
where the vector $\vec{z}=(0,0,1)$. For $\Delta_0=-g(2N+3)$ as in Fig.~\ref{fig:floquet}, we extract the norm of $H_{(2)}/N$ in leading order of $N$ as,
\begin{equation}
    \frac{\norm{H_{(2)}}}{N} \sim \frac{10 cg\Delta_0 }{3\omega_d}\approx \frac{20 cg^2N}{3\omega_d},
\end{equation}
where $c$ is an $O(1)$ factor depending on the dimensionless norms $\norm{\vec{r}_{l,\pm}\times((\vec{r}_{j,\pm})^*\times\hat{z})}/g\omega_d$. The condition $\norm{H_{(2)}}\ll\norm{H_{(1)}}$ then simplifies to the following $N-$dependent criteria:
\begin{equation}
    ||H_{(1)}||
    \sim
    3gN \gg ||H_{(2)}|| \sim \frac{20g^2 N^2}{3\omega_d}
    \quad \implies \quad \omega_d\gg\tfrac{20}{9}gNA
\end{equation}
Further, for the rotating wave approximation to hold, we need $N\omega_0$ to be the largest energy scale in the Floquet Hamiltonian $H^F$, i.e. much greater than $||H_{(1)}||$. This translates to the condition,
\begin{equation}
    \omega_0\gg g 
\end{equation}
For $\Delta_0 = -g(2N+3)$ (i.e. $\Delta=0$), we obtain the simple criteria $\omega_0\gg g$, which is independent of $N$. Therefore, to generate the target Hamiltonian for a higher photon number, the driving frequency should increase linearly with $N$. In Fig.~\ref{fig:floquet}, we show numerical plot of the stroboscopic cavity populations in the lab frame by time-evolving with the Hamiltonian Eq.~\eqref{eq:high_freq}. We see that after long stroboscopic times, the agreement between the Floquet-generated Hamiltonian and the static one gets worse. This is because of the contribution of higher order terms in the high-frequency expansion. We checked that the total number of photons in the rotating frame is constant for all times shown in Fig.~\ref{fig:floquet}.  \\

\section{External driving and non-reciprocal routing of power}
In this section, we investigate the three cavity-one qubit device coupled to an external drive and detection ports. We find, as expected, that the device can be used as a unidirectional router from cavity 3 to cavities 1 or 2, depending on the sign of the Chern number.

The specific setup is as follows. We drive one of the cavities (cavity $3$) with an external drive close to its resonance frequency $\omega$ to initialize it in a desired state. The remaining two cavities are coupled to two external {\it{detector}} cavities ($D_1$ and $D_2$) relatively strongly and in a non-reciprocal way. Each such detector cavity mimics the effect of an impedance matched transmission line; the power in cavity $2$ ($1$) leaks into output cavity $D_2$ ($D_1)$, with minimal reflection. 

The Hamiltonian of the full setup is
\begin{equation}
    H(t) = F^*(t)b_3+F(t)b_3^{\dagger}+r_{out}b^{1\dagger}_{out}b_1 + r_{in}b^{\dagger}_1b_{out}^1 +  r_{out}b^{2\dagger}_{out}b_2 + r_{in}b^{\dagger}_2b_{out}^2+\omega \hat{N} + H,
\end{equation}
where $b^{1}_{out}$ ($b^{2}_{out}$) are modes of the output cavities $D_1$ ($D_2$), $H$ is the Hamiltonian in main text,
\begin{equation}
H = \sum\limits_{j}b_j^{\dagger}b_{j+1}G_{j-1}+h.c
\end{equation}
and,
\begin{equation}
    F(t) = \frac{1}{2\pi}\int_{-\infty}^{\infty}F_0 e^{i\omega' t}e^{-\sigma|\omega'-\omega|}d\omega'.
\end{equation}
In the limit $\sigma\rightarrow 0$, this produces a delta function kick at time $t=0$. For a finite $\sigma$, the amplitude is modulated by a Lorentzian pulse centred at the resonance frequency $\omega$,
\begin{equation}
    F(t) = \frac{F_0\sigma e^{i\omega t}}{\pi(t^2+\sigma^2)}.
\end{equation}
We start from the vacuum state of the cavities $\ket{\psi(0)}=\ket{0,0,0}\ket{+}$. Recall that $T=4\pi/g\sqrt{3}$ is the period of circulation for the isolated setup.

We choose (i) the ratio of the coupling constants to the detector cavities $r_{out}/r_{in}\gg 1$ to mimic the coupling to a transmission line, and (2) $2 \pi r_{out}^{-1} \ll T$ so that power in cavities 1 and 2 leaks out on a faster time scale as compared to the period of circulation, and (3) $\sigma \ll T$ to initialize cavity 3 before the state begins to circulate. Note that $r_{out}/r_{in}\neq 1$ makes the Hamiltonian non-hermitian. 

First, let $r_{out}=r_{in}=0$. Since the duration of the drive is much smaller than the period of circulation, we can effectively decouple the action of the drive from the circulating dynamics. The drive prepares the driven cavity in a coherent state with the mean number of photons $|F_0|^2/4$. 

With non-zero coupling to the output detectors (such that $r_{out}/r_{in}\gg 1$), the power in cavity $3$ flows into either one of the two detectors, depending on the direction of circulation. To quantify the detected signal through $D_1$ and $D_2$, we calculate the imbalance,
\begin{equation}
    \mathcal{I}(t) = \frac{\braket{b^{1\dagger}_{out}b^{1}_{out}}_t-\braket{b^{2\dagger}_{out}b^{2}_{out}}_t}{\braket{b^{1\dagger}_{out}b^{1}_{out}}_t+\braket{b^{2\dagger}_{out}b^{2}_{out}}_t},
\end{equation}
where the expectation values are taken with respect to the time dependent state, i.e,  $\braket{b^{1\dagger}_{out}b^{1}_{out}}_t=\braket{\psi(t)|b^{1\dagger}_{out}b^{1}_{out}|\psi(t)}$. Depending on whether the photons flow into the detectors $1$ or $2$, $\mathcal{I}(t)$ reaches either $+1$ or $-1$, respectively in the steady state. \\
\begin{figure*}
	\includegraphics[width=\textwidth,height=4.5cm]{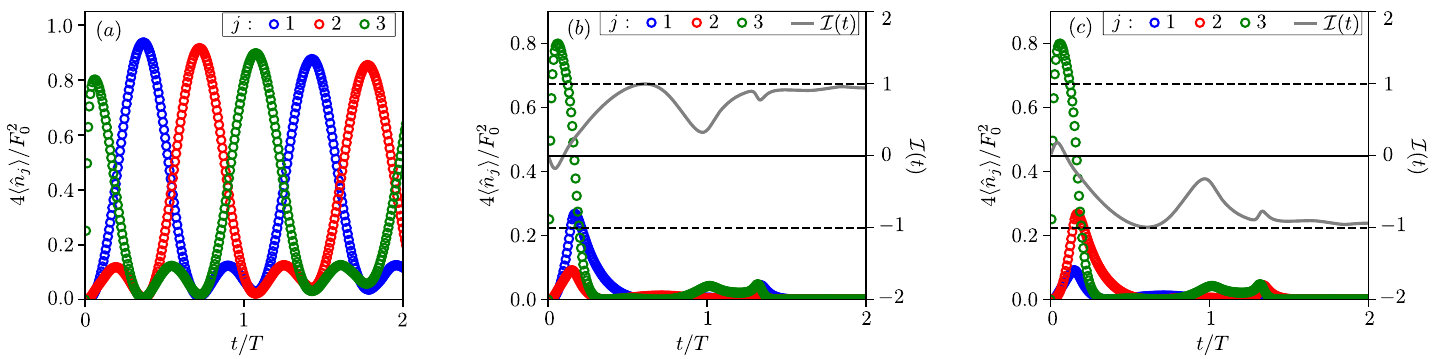}
			
	\caption{(a) Unitary circulation starting from the vacuum state $\ket{0,0,0}\ket{+}$ with a drive attached to cavity $3$ without coupling to output cavities. (b), (c) With finite coupling to the output cavities, output power is collected in either of the two detectors $D_1$ or $D_2$, depending on the direction of circulation. Parameters: $g=0.5$ in (b) and $g=-0.5$ in (c), $\omega=1.0$, $F_0=5.0$, $\sigma=0.1$, $r_{out}=2.0$ and $r_{in}=0.02$.}
	\label{fig:drive_detect}
\end{figure*}
Fig.~\ref{fig:drive_detect} shows simulation results in the driven setup without (panel (a)) and with (panels (b,c)) detectors. Fig.~\ref{fig:drive_detect}(a) shows the cavity populations of the three cavity-one qubit device starting from the initial state $\ket{0,0,0}\ket{+}$. We see that the drive successfully prepares an initial state with the desired number of average photons, and this population subsequently circulates. Fig.~\ref{fig:drive_detect}(b,c) plot the cavity populations and the imbalance when the detectors are coupled. In (b), we see that the photons flow from cavity $3$ to $1$, and then leak out into detector $D_1$. The imbalance thus reaches a value close to $+1$ in a couple of periods. Fig.~\ref{fig:drive_detect}(c) shows that $\mathcal{I}(t)$ reaches a value close to $-1$ when the sign of the coupling $g$ is flipped. This is as expected from the reversal of the direction of circulation in the closed system.